\def\year{2021}\relax
\documentclass[letterpaper]{article} 
\usepackage{aaai21} 
\usepackage{times} 
\usepackage{helvet} 
\usepackage{courier}  
\usepackage[hyphens]{url} 
\usepackage{graphicx} 
\usepackage{natbib}  
\usepackage{caption} 
\nocopyright
\title{RAWLSNET: Altering Bayesian Networks to Encode \\ Rawlsian Fair Equality of Opportunity}
\author{
   David Liu$^{*\dag}$, Zohair Shafi$^{*\dag}$, Will Fleisher$^\dag$, Tina Eliassi-Rad\thanks{The first two authors contributed equally to the paper. The point-of-contact is eliassi@northeastern.edu.}$^\dag$, Scott Alfeld$^\ddag$
    \\
}
\affiliations{
    $^\dag$Northeastern University, $^\ddag$Amherst College \\
    
}

\usepackage{url}
\usepackage{verbatim}
\usepackage{caption}
\usepackage{subcaption}
\usepackage[ruled,vlined]{algorithm2e}
\usepackage{mathtools,amsmath}
\usepackage{graphicx}
\usepackage{xcolor}
\newcommand{\queryvar}{Q}
\newcommand{\controlvar}{C}

\newcommand{\query}{q}
\newcommand{\control}{c}

\newcommand{\parents}[1]{{\texttt{Par}\left(#1\right)}}

\newcommand{\tmpvar}[2]{Z^{(#1)}_{#2}}

\newcommand{\normconstant}{\alpha}
\newcommand{\altnormconstant}{\tilde \normconstant}

\newcommand{\talent}{T}
\newcommand{\ses}{SES}
\newcommand{\exam}{E}
\newcommand{\college}{C}
\newcommand{\goodjob}{J}

\newcommand{\allvars}{{\pmb V}}

\newcommand{\justifiedvar}{J}
\newcommand{\justifiedvars}{{\pmb \justifiedvar}}
\newcommand{\justifiedval}{j}
\newcommand{\justifiedassigns}{{\pmb \justifiedval}}

\newcommand{\sensitivevar}{S}
\newcommand{\sensitivevars}{{\pmb \sensitivevar}}
\newcommand{\sensitiveval}{s}
\newcommand{\sensitiveassigns}{{\pmb \sensitiveval}}

\newcommand{\othervar}{O}
\newcommand{\othervars}{{\pmb \othervar}}
\newcommand{\otherval}{o}
\newcommand{\otherassigns}{{\pmb \otherval}}

\newcommand{\indep}{\perp \!\!\! \perp}

\newcommand{\someassign}{{\pmb a}}

\newcommand{\RAWLSNET}{{\texttt{RAWLSNET}}}
\usepackage{algorithmic}

\newcommand{\hide}[1]{}

\begin{document}

\maketitle

\begin{abstract}
We present \RAWLSNET, a system for altering Bayesian Network (BN) models to satisfy the Rawlsian principle of \textit{fair equality of opportunity} (FEO).  \RAWLSNET's BN models generate aspirational data distributions: data generated to reflect an ideally fair, FEO-satisfying society.
FEO states that everyone with the same talent and willingness to use it should have the same chance of achieving advantageous social positions (e.g.,  employment), regardless of their background circumstances (e.g., socioeconomic status). Satisfying FEO requires alterations to social structures such as school assignments. Our paper describes \RAWLSNET, a method which takes as input a BN representation of an FEO application and alters the BN's parameters so as to satisfy FEO when possible, and minimize deviation from FEO otherwise. We also offer guidance for applying \RAWLSNET, including on recognizing proper applications of FEO. We demonstrate the use of our system with publicly available data sets. \RAWLSNET's altered BNs offer the novel capability of generating \textit{aspirational data} for FEO-relevant tasks. Aspirational data are free from the biases of real-world data, and thus are useful for recognizing and detecting sources of unfairness in machine learning algorithms besides biased data. 

\end{abstract}

\begin{figure*}
    \centering
    \includegraphics[scale=0.45]{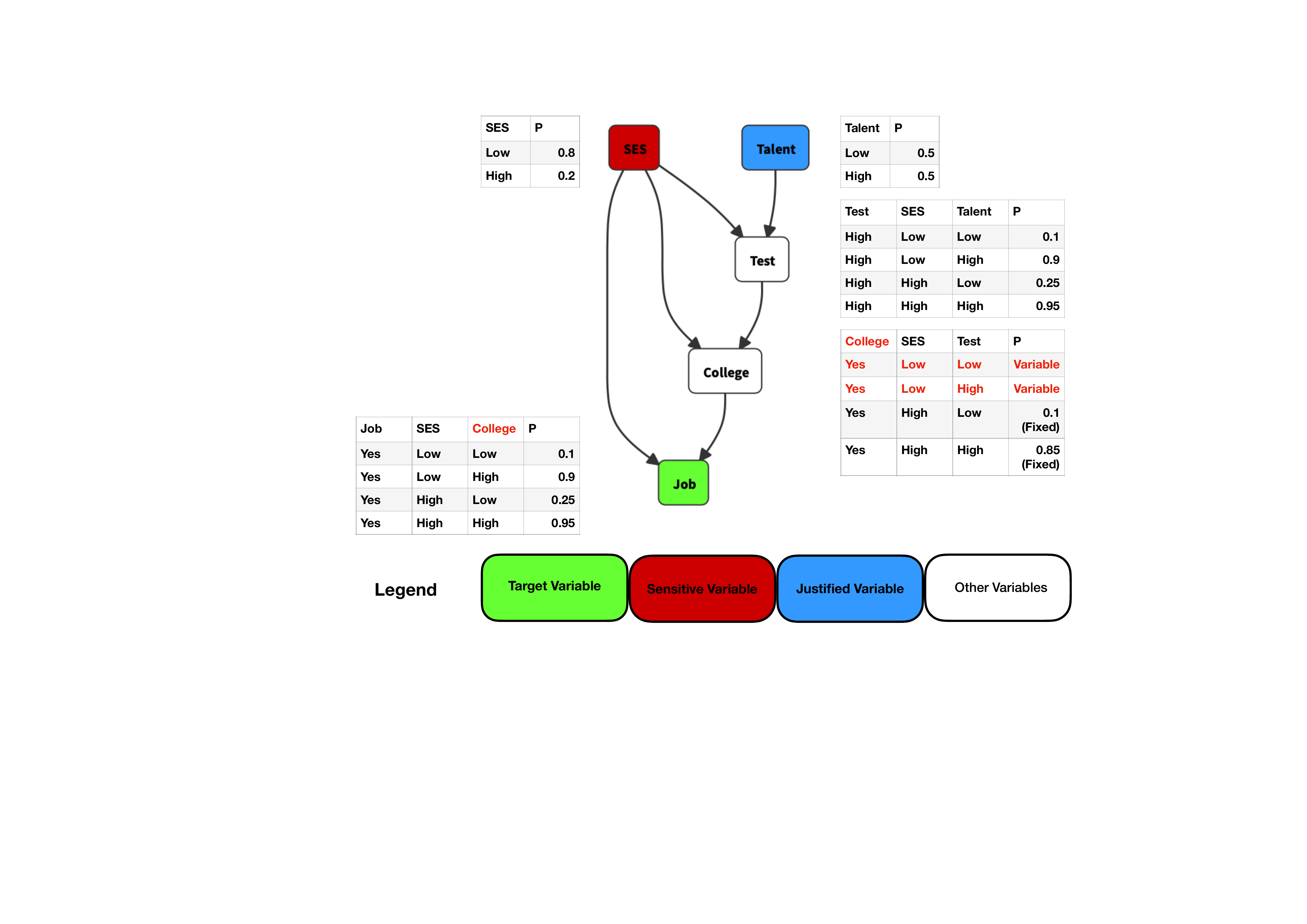}
    \caption{Running example of a Bayesian network for college admissions. The Talent variable refers to having the innate capability to succeed at a job. The Socioeconomic Status (SES) node represents socioeconomic status (e.g., an individual with a high SES is often wealthy). Test Score refers to GPA and entrance exam results. The College variable stands for admissions into college and is highlighted in the conditional probability tables (CPTs) to indicate that it  represents the decision being modeled. Good Job refers to whether the individual attained desirable employment. Note: For brevity, the CPTs for Test, College, and Job only contain the probabilities for when these values are one.}
    \label{fig:collegeBN}
\end{figure*}

\section{Introduction} \label{sec:intro}
Machine learning algorithms often display pernicious biases that lead to harmful and unfair outcomes for marginalized groups \cite{oneil, broussard, eubanks, fry, noble, benjamin}. In response to this algorithmic bias, there is a widely growing literature seeking to achieve fairness in machine learning \cite{barocas-hardt-narayanan,kearns:roth,pessach2020algorithmic}.

As a guide to fairness, we appeal to John Rawls' theory of \emph{justice as fairness}~\cite{rawls1971,rawlsTheoryJusticeRev1999}. 
For a society to be fair in the Rawlsian sense, it must satisfy a substantive equality of opportunity principle, which Rawls calls \textit{fair equality of opportunity} (FEO). This principle governs fair decision-making in the context of distributing desirable social positions (such as employment) in society. Specifically, it requires that all people who have the same level of talent and ambition have the same chance of attaining advantageous social positions. 

Here, we address the following question:
\emph{Given an unfair (in the Rawlsian sense) outcome and the capability to alter some (but not all) decision-making  processes, how can one satisfy FEO?} To this end, we use Bayesian Networks (BNs) to model decisions that are governed by FEO --- namely, those that impact the distribution of advantageous social positions. 
We then give a characterization of FEO in terms of conditional probability, which allows for the mathematical formalization of the above problem.

As a solution to this problem, we present \RAWLSNET. Given data, \RAWLSNET\ offers guidance on making decisions that satisfy FEO.  When satisfying FEO is impossible (e.g., due to resource constraints such as number of available jobs), \RAWLSNET\ finds a ``closest'' solution  to satisfying FEO. See Section~\ref{sec:approach} for details on ``closest.''

\RAWLSNET\ has the following three components: (1) learn a BN; (2) determine relevance to FEO; (3) update parameters of the learned BN to satisfy FEO if possible. Otherwise, update the parameters to approximately satisfy FEO. To learn the BN, \RAWLSNET\ accepts as input (\emph{i}) a fully trained Bayesian Network (BN), or (\emph{ii}) the structure of a BN and data to learn the BN's parameters (i.e., probabilities in its conditional probability tables -- a.k.a.~CPTs), or (\emph{iii}) data to learn the BN's structure and its parameters. If \RAWLSNET\ has to learn the BN structure, it asks the user to answer a series of questions:

\begin{itemize}
\item Identify variables that are morally justified for unequal decisions (e.g., talent)
\item Identify variables that are sensitive and not justified for unequal decisions (e.g., gender, socioeconomic status, race, etc).
\item Identify variables over which you have control (e.g., college admissions).
\item Identify the variable that can be considered as a socially advantageous position (e.g., job)
\item Identify variables in your data which you believe RAWLSNET should ignore.
\end{itemize}

To illustrate the method and purpose of \RAWLSNET, we present a running example and discuss two functions of \RAWLSNET: generating aspirational data and providing policy-advice to domain experts.

\subsubsection{A running example.}
To illustrate  FEO and \RAWLSNET\, we use the following running example in the domain of education. We consider people applying for a job, where each applicant is born with a certain degree of talent and a certain socioeconomic status (SES). 
Subsequently, applicants achieve a certain secondary GPA and take college entrance exams.
On the basis of these exam scores (including GPA) and their SES, applicants have a certain probability of getting into a prestigious college.
After graduation, each applicant's college experience and SES determine how likely it is that he/she obtains a good job.
We imagine that the initial situation, prior to \RAWLSNET's intervention, reflects certain unjustified advantages for those with high SES. As is plausibly true for modern societies, high SES has an impact on test scores, college acceptance, and employment.
Most relevantly, applicants with high SES have a better chance of achieving a good job than similarly talented people with low SES.
This situation is a violation of FEO. 
In order to satisfy FEO,  whether or not an applicant is given a job must be independent of the applicant's SES, given their talent.
For exposition, we model all variables as binary and include only a single sensitive feature (which is SES), but note that neither of these limitations are necessary. Figure~\ref{fig:collegeBN} shows the BN for this example.

We consider the task of a hypothetical college admissions committee as they decide who is/isn't accepted into college based on the test scores and SES of potential applicants.
Their goal is to achieve FEO, which we note is a consideration at the hiring level, not the college admittance level.
By altering the probabilities of accepting low-SES applicants (based on their test scores), the college admissions committee can provide an advantage to low-SES applicants.
This advantage can counteract the unfair benefits of high-SES students that directly influence both their test scores and job placement. 
\RAWLSNET\ sets the probabilities $P(College | SES, Test)$ such that FEO is satisfied.

\subsubsection{Aspirational data.}
\RAWLSNET\, can be used to generate \textit{aspirational data}: synthetic data that models ideally fair circumstances.
Aspirational data can be sampled from the FEO-satisfying BN models generated by \RAWLSNET. 
This data can then be used to aid in research on algorithmic bias by shifting the focus towards sources of bias introduced downstream in the machine learning life cycle, such as training and deployment processes.  
By evaluating data analysis methods on aspirational data, one can determine whether or not ``bad data'' is the only factor to blame in a biased system.

Complementary to our method of generating aspirational data are methods of debiasing or cleaning data \cite{BeutelData2017,boukbasiAdvances2016,HeGeometric2020}.
Importantly, however, \RAWLSNET\ acts on the \emph{distribution} from which the data comes, rather than the data itself.
This yields two key advantages over debiasing methods.
First, the resulting BN can be sampled from to generate aspirational datasets of any size and thus used to run a wider variety of experimental investigation than what any single debiased dataset could offer. 
Second, the alterations to the BN --- which we note are changes to the conditional probabilities and not the structure of the BN --- can be directly interpreted as policy advice, as we discuss next.

\subsubsection{Policy advice.}
Another use of \RAWLSNET\ is to aid in decision-making for policy makers.
If a user knows the distribution of talent for their task, then \RAWLSNET, can be used to inform their decisions.
This knowledge may be learned using a method to infer hidden variables from data (e.g., as in \citealt{ChenFairness2019}).
Alternatively, the user may appeal to domain experts with knowledge about job-relevant talent.
For instance, \RAWLSNET\ might be used to advise acceptance decisions for a college admissions committee.
The committee may have domain experts with knowledge of the distribution of talent in the applicant pool, or  have adequate data to infer the distribution of talent as a hidden variable.
Assuming distribution of talent is available in one of these ways, \RAWLSNET\ will calculate the acceptance rates for applicants from different groups needed to satisfy FEO.

\subsubsection{Contributions.}
We introduce \RAWLSNET, a method for altering a BN so as to satisfy Rawlsian FEO. To our knowledge, \RAWLSNET\ is the first tool developed that produces aspirational (FEO) data distributions. \RAWLSNET\ alters BN models in a way that preserves their initial structure. This preservation allows it to generate aspirational data for important social systems. Whenever satisfying FEO is not possible, \RAWLSNET\ outputs a BN whose distribution is closest to satisfying FEO.

\subsubsection{Paper structure.}
We provide background on FEO and its appropriate applications in the next section. Then, we discuss the formalization of our problem and present our contribution: \RAWLSNET\ and its underlying mathematics. This is followed by experiments, related work, and discussion.

\section{Background: Fair Equality of Opportunity} \label{sec:background}
FEO requires that any two people with similar talent and ambition receive the same chance of achieving an advantageous social position (e.g., a good job), regardless of their background. This principle is designed to eliminate the effects of discrimination and other oppressive structures in determining who has access to advantageous social positions. It requires that social features irrelevant to determining who will do best at a job are irrelevant to determining who will receive the job.

FEO is one aspect of the Rawlsian theory \textit{justice as fairness} \cite{rawls1971,rawlsJusticeFairnessRestatement2001}. This theory is influential, well-supported, and widely popular \cite{Brighouse2005-BRIJ}. FEO itself is particularly plausible, as similar principles appear in a variety of other theories \cite{sep-equal-opportunity}. Moreover, FEO is formally similar to other kinds of substantive equality of opportunity principles \cite{heidariMoralFrameworkUnderstanding2019,LEFRANC20091189,roemerEqualityOpportunity2009a}. Thus, our work here can be applied to satisfying those principles as well.

\citet{rawls1971,rawlsTheoryJusticeRev1999,rawlsJusticeFairnessRestatement2001} offered a theory of what is required for a society to be ideally fair. Rawls' theory of fairness requires that any wealth inequalities in society must be attached to advantageous social positions, what Rawls called \textit{offices}. An office is desirable employment that carries with it greater responsibility, greater prestige, and/or higher pay. Rawls' theory requires that advantageous positions must be open to all applicants under FEO. Thus, FEO applies directly to decisions about employment. Obtaining an office is the only way, in a fair society, to become wealthier than your peers. In sum, FEO governs how good jobs are handed out.

Rawls called the sort of equality of opportunity we are interested in ``fair'' to contrast it with a more familiar sort, which is often called \textit{formal equality of opportunity} (or formal EO for short).
Formal EO requires two things: (1) that positions of social advantage be open to all applicants, and (2) that applicants be evaluated entirely based on their qualifications for the position \cite{rawlsJusticeFairnessRestatement2001,Arneson1999-ARNARE}. Formal EO rules out explicit discrimination based on group membership (e.g., race and gender). It also rules out caste systems, nepotism, and favoritism. Formal EO is essentially an ideal version of what contemporary equality of opportunity laws governing housing and employment aim at. While itself quite stringent, formal EO is compatible with a wide variety of oppressive structures and implicit discrimination. For instance, the hiring process for an engineering position might conform to formal EO if it widely publicizes its openings, considers all applications, and hires the most qualified engineers. But if only men are allowed to attend engineering schools, the result is still unfair. Formal EO is an important principle of fairness. \RAWLSNET\ is designed to satisfy FEO without creating violations of Formal EO. It achieves this by altering the probabilities that affect obtaining advantageous positions. 

The primary difference between formal EO and FEO concerns which features of an applicant are relevant to justifying hiring choices. Formal EO focuses on qualifications: the skills, training, and experience that an applicant has \textit{at the time of hiring}, which determine how good the applicant would be at the job. In a real-world, contemporary society, sensitive but morally irrelevant features can make a significant difference to what qualifications an applicant has. Being born into a high SES family, for instance, has a large impact on the kind of education one has access to. In the imagined (but realistic) engineer case mentioned above, it was gender that was inappropriately affecting educational opportunities. It is the focus on qualifications that makes formal EO inadequate to fully ensure genuine fairness. Yet qualifications are clearly enormously important: we do not want our bridges built by engineers who did not go to engineering school. 

FEO is a principle meant to fill the gap between what formal EO requires and what is required for a genuinely fair EO (hence the name). It focuses not on qualifications at the time of hire, but instead on innate \textit{talent} and \textit{ambition}. Here, we can understand talent to be an innate potential to be good at some job. Ambition here is one's willingness to develop and use his/her talent. For most of us, no amount of training, experience, and hard work could make us into basketball players as good as Lebron James. What he has, in addition to an incredible work ethic and ambition, is innate talent most of us lack. Similarly, there are many other innate features which effect an individual's potential to excel at various jobs. Following Rawls, we label these features, generically, as \textit{talent}. A person's willingness to work to develop and employ his/her talents constitutes his/her ambition.

According to Rawls, FEO requires that ``those who are at the same level of talent and ability, and have the same willingness to use them, should have the same prospects of success regardless of their initial place in the system'' \cite[p.~63]{rawlsTheoryJusticeRev1999}. Two people with the same talent and degree of ambition should have equal chances of obtaining desirable employment, and the social benefits it brings with it. In other words, one's chance of getting a good job should be statistically independent of any features of an individual other than their talent and ambition.\footnote{%
    Strictly speaking, what is required is that any two applicants with the same talent and ambition should have the same probability of getting the job \textit{when not conditioned on other attributes such as training}.
}
FEO requires that such features as ethnicity, gender, LGBTQ+ status, nationality, birthplace, etc., must all be statistically irrelevant to whether one achieves an office. Only talent and ambition should ultimately make a difference to whether you get the job. This is not because talent is relevant to determining who deserves social advantages. Rather, it is because everyone is made better off if people are incentivized to develop and employ their talents. The benefits of this incentivization are what justify any inequality in the first place. 

An extremely plausible (and perhaps morally obligatory) assumption is that talent is distributed independently of social group features such as race, ethnicity, gender, LGTBQ+ status, etc. We call these \textit{sensitive features}. On this assumption, differences in qualifications correlated with these sensitive features must be the result of differences in experience and training. FEO requires that such differences be made ultimately irrelevant to determining who receives an advantageous position. Any two applicants with the same talent and ambition should have the same probability of obtaining desirable employment. Satisfying FEO therefore requires removing or ameliorating the impact of sensitive features on employment. 

Crucially, FEO must be satisfied in a way that avoids violating \textit{formal} EO. Satisfying both principles requires changing the way talent and ambition are related to qualifications.\footnote{%
    There is significant dispute about the relationship between formal and fair equality of opportunity. In particular, there is dispute about whether satisfying FEO requires also satisfying formal EO \cite{Arneson1999-ARNARE,Taylor2004-TAYSAT-3,Taylor2009-TAYRAA}. We seek to sidestep this dispute. We take both principles to be important, and remain neutral on their relative priority.
} 
\RAWLSNET\, is designed to allow a decision-maker to satisfy FEO (when possible) without creating violations of formal EO. It accomplishes this because it is designed to operate on decisions that are made \textit{prior to the point of hiring for advantageous positions}. Such decisions are not directly governed by formal EO, as that principle concerns employment decisions exclusively. Our strategy is illustrated by our college admissions example. There, the decision in question is whether to admit someone to a prestigious college. But being a student at such an institution is not an office: it is not a job by which unequal wealth is distributed. Admissions do indirectly impact social advantage, but only insofar as they impact hiring.

\section{Proposed Approach: \RAWLSNET}
\label{sec:approach}

In this section we define what it is to be an {\it FEO Application}. We also provide guidance for determining whether a task is an FEO application.
We then present \RAWLSNET, a method for altering the parameters of a BN model for an FEO application in order to satisfy FEO.
We also discuss runtime considerations and encoding constraints of the underlying application.

\subsection{FEO Applications}
FEO governs the distribution of advantageous social positions. Applied to our running example involving college admissions, \RAWLSNET\, is designed to determine the correct college acceptance rates (i.e., the probability of being admitted) to ensure that FEO is satisfied at the later stage of hiring. Our running example is what we call an \textit{FEO application}. ``FEO application'' is a novel term which we define as: \textit{a decision which affects whether FEO is satisfied by a distinct, subsequent hiring decision}. Genuine FEO applications are decisions that are needed to satisfy FEO, but which do not introduce violations of formal equality of opportunity. We avoid violations of formal EO by using earlier decisions to improve the qualifications of talented applicants prior to the hiring process. In the college admissions example, this involves making it more likely that talented students from low SES backgrounds are admitted to college, which improves their qualifications for being hired. Thus, in our running example we satisfy FEO without violating formal EO by intervening on the earlier admissions decision.

\subsubsection{A Guide for Recognizing FEO Applications.}
For a decision to count as an FEO application, it must meet three conditions: it must (1) affect the distribution of advantageous social positions (i.e., good jobs), (2) be made prior to hiring decisions, and (3) be made on the basis of appropriate features of applicants.
This is illustrated by our college admissions example, which meets all of these conditions.
These are \textit{necessary} conditions for being an FEO application.
Moreover, any decision that meets these conditions is likely to be such an application.
Regarding condition (1), recall that offices are desirable employment positions that carry with them the social advantages that lead to inequality.
Condition (2) helps ensure that the decision-maker does not introduce violations of \textit{formal} EO.
 Condition (3) concerns the relevant features of applicants. In an FEO application, we distinguish three categories of relevant features. There are \textit{justifying} features: features that morally justify the inequalities which are attached to advantageous positions. Following Rawls, these are limited to talent(s) and ambition. There are also \textit{sensitive} features: features which should not be allowed to lead to inequalities. As mentioned, these are features such as race, gender, and socioeconomic status. Finally, there are what we simply call ``other'' features. Other features may permissibly impact hiring and other decisions, but which do not themselves justify inequality. For the purposes of recognizing an FEO application, its crucial to identify the relevant justifying features and sensitive features. 

In our college admissions example, the \textit{other} category includes the exam score feature and the college admissions decision itself. In fact, this is an important feature of an FEO application decision: it will always concern a feature that is neither sensitive nor justifying. This is because, as we have defined the term, an FEO application is always a decision by someone other than the applicant, and is one that indirectly impacts the distribution of advantageous positions.

\subsection{Altering Bayesian Networks with \RAWLSNET}
\label{sec:notation}

\RAWLSNET\, provides a method for altering a given BN (describing an FEO application) such that its defined distribution satisfies the Rawlsian FEO principle. \RAWLSNET\ can work when data is provided with or without a BN. In the latter case, it will learn a BN structure that best fits the data. In the discussion to follow, we assume a BN structure has been given along with the data with variables \(\mathcal V\), edges \(\mathcal E\), and parameters \(\mathcal P\) -- i.e., elements of the conditional probability tables (CPTs) of each variable.  The variables are partitioned as \(\mathcal V = \justifiedvars \cup \sensitivevars \cup \othervars\) where \(\justifiedvars\) is the set of variables morally justified for inequality (e.g., talent), \(\sensitivevars\) is the set of sensitive variables (e.g., socioeconomic status), and \(\othervars\) is the set of remaining (other) variables. In addition, a {\it control} variable \(\controlvar \in \othervars\) is specified as the variable which we can control the CPT of (e.g., college admissions) and a {\it target} variable \(\queryvar \in \othervars\) (e.g., obtaining a good job). One can optionally supply \RAWLSNET\, a set of {\it feasibility constraints}, which we discuss below. Table~\ref{table:notation} summarizes our notation.

\begin{table*}[!ht]
  \centering
  \begin{tabular}{|c|c|}
    \hline
    Symbol & Meaning \\ \hline
\(\justifiedvars = \justifiedvar_1, \ldots, \justifiedvar_{|\justifiedvars|}\) & The set of justified variables       \\
\(\sensitivevars = \sensitivevar_1, \ldots, \sensitivevar_{|\sensitivevars|}\) & The set of sensitive variables       \\
\(\othervars = \othervar_1, \ldots, \othervar_{|\othervars|}\) & The set of other variables        \\
\(\pmb v = v_1, \ldots, v_{|\pmb V|}\) & An assignment to the variables in the corresponding set \(\pmb V\) \\
\(\parents{V; \justifiedassigns, \sensitiveassigns, \otherassigns}\) & The set of variables that are parents of \(V\), and their assignments in \(\justifiedassigns, \sensitiveassigns, \otherassigns\)\\
\(\queryvar, \query\) & The target variable, a particular assignment of it \\
\(\controlvar, \control\) & The control variable, a particular assignment of it\\
\hline
  \end{tabular}
  \caption{ Notation table.
    \RAWLSNET~ is provided a BN with variables \(\justifiedvars \cup \sensitivevars \cup \othervars\).
    In addition, a control variable \(\controlvar \in \othervars\) and target variable \(\queryvar \in \othervars\) are specified.
    The output of \RAWLSNET~ is a new BN, identical to the original in structure and all parameters except select elements of the CPT for \(\controlvar\), which are edited such that Rawlsian FEO is satisfied for the target variable \(\queryvar\).}
  \label{table:notation}
  \end{table*}

In this context, FEO is obtained by ensuring statistical independence between the target variable $Q$, and the sensitive variables \sensitivevars, conditioned on the justified variables \justifiedvars.
That is, we seek to set the CPT values of the control variable \(\controlvar\) such that:
\begin{align}
\queryvar \indep \sensitivevars ~|~ \justifiedvars
\end{align}
Equivalently, we seek values for the conditional probability of the control variable $\controlvar$ given its parents such that: %\(P(\controlvar | \parentvar_1 \ldots \parentvar_k)\) such that:
\begin{align}
P\left(\queryvar ~|~ \justifiedassigns\right) = P\left(\queryvar ~|~ \justifiedassigns \cup \sensitiveassigns \right) \label{eqn:set_of_goals}
\end{align}
for every possible assignment \(\justifiedassigns\) of variables in \(\justifiedvars\)  and assignment \(\sensitiveassigns\) of variables in \(\sensitivevars\).
For any particular assignments \(\justifiedassigns, \sensitiveassigns, \otherassigns\) of justified, sensitive, and other variables respectively, we let:
\begin{align}
  \parents{V; \justifiedassigns, \sensitiveassigns, \otherassigns}
\end{align}
denote the set of variables (and their assignments in \(\justifiedassigns, \sensitiveassigns, \otherassigns\)) that are the parents of \(V\).

Note that (\ref{eqn:set_of_goals}) is a collection of desired qualities.
For ease of notation, we focus on just one.
Namely, our goal is to satisfy:
\begin{align}
P\left(\query ~|~ \justifiedassigns \right) = P\left(\query ~|~ \justifiedassigns, \sensitiveassigns\right) \label{eqn:goal}
\end{align}
We first note that:
\begin{align}
  P\left(\query ~|~ \justifiedassigns \right) &= \normconstant P\left(\query, \justifiedassigns \right) \label{eqn:deriv_start}\\
  &= \normconstant \sum_{\tilde \otherassigns} \sum_{\tilde \sensitiveassigns} P\left(\query, \justifiedassigns, \tilde \otherassigns, \tilde \sensitiveassigns \right)
\end{align}
where \(\normconstant\) is a normalization constant, and \(\tilde \otherassigns\) ranges over all possible assignments for \(\othervars\) (resp. \(\tilde \sensitiveassigns\), \(\sensitivevars\)).
We let  \(\tmpvar{q}{\justifiedassigns, \tilde \otherassigns, \tilde \sensitiveassigns} = P\left(q ~|~ \parents{Q; \justifiedassigns, \tilde \otherassigns, \tilde \sensitiveassigns}\right)\) and for any set \(\allvars\) of variables and assignment \(\someassign\) of its parents we let:
\begin{align}
  \tmpvar{\allvars}{\someassign} &= \prod_i P(V_i ~|~ \parents{V_i; \someassign})
\end{align}
For ease of notation, we suppress the dependency on the assignment when clear  and simply write \(\tmpvar{q}{}, \tmpvar{\pmb S}{}\).
With this notation we have: 
\begin{align}
  P\left(\query ~|~ \justifiedassigns \right) &= \sum_{\tilde \otherassigns} \sum_{\tilde \sensitiveassigns}\tmpvar{q}{} \tmpvar{J}{} \tmpvar{S}{} \tmpvar{O}{}
\end{align}
Separating out the value in \(\tmpvar{O}{}\) corresponding to \(\controlvar\) we let:
\begin{align}
\tmpvar{O}{\justifiedassigns, \tilde \otherassigns, \tilde \sensitiveassigns} &= \tmpvar{O'}{\justifiedassigns, \tilde \otherassigns, \tilde \sensitiveassigns} P\left(\control | \parents{\controlvar; \justifiedassigns, \tilde \otherassigns, \tilde \sensitiveassigns}\right)
\end{align}
to obtain:
\begin{align}
P\left(\query ~|~ \justifiedassigns \right) =  \sum_{\tilde \otherassigns} \sum_{\tilde \sensitiveassigns}\tmpvar{q}{} \tmpvar{J}{} \tmpvar{S}{} \tmpvar{O'}{}P\left(\control | \parents{\controlvar; \justifiedassigns, \tilde \otherassigns, \tilde \sensitiveassigns}\right) \label{eqn:lhs}
\end{align}
Similarly, for the right-hand side of (\ref{eqn:goal})  we have:
\begin{align}
 P\left(\query ~|~ \justifiedassigns, \sensitiveassigns\right) &= \altnormconstant
\sum_{\tilde \otherassigns} \tmpvar{q}{} \tmpvar{J}{} \tmpvar{S}{} \tmpvar{O'}{}P\left(\control | \parents{\controlvar; \justifiedassigns, \tilde \otherassigns, \tilde \sensitiveassigns}\right) \label{eqn:rhs}
\end{align}
Therefore we can satisfy Rawlsian FEO by satisfying the equality (\ref{eqn:lhs})~=~(\ref{eqn:rhs}) for each assignments \(\justifiedassigns\) and \(\sensitiveassigns\). This yields a system of linear equations, solvable using standard methods.

\subsubsection{Feasibility Constraints.}
\label{sec:feasibilityconstraints}

Consider our running example of college admissions.
Using \RAWLSNET, one can select an admission policy so as to ensure fair (in the Rawlsian sense) job allocation.
However, if we solve the system of equations described above, there is no guarantee that the resulting admission policy is satisifiable in practice --- it may require that the school admits substantially more (or fewer) students than is viable.
We therefore imbue \RAWLSNET\, with the capability to accept a set of constraints.

Given an equality constraint on some marginal distribution of the BN (e.g., the expected number of students admitted to college is precisely \(p\) percent of the population), we simply add this to the collection of equations defined above.
With similar derivation to (\ref{eqn:deriv_start}) through (\ref{eqn:lhs}), we note that the above constraint is also linear and thus the system is still efficiently solvable.
Given inequality constraints (e.g., a particular marginal probability must be in some interval), \RAWLSNET\, solves a linear program.

\subsubsection{Runtime.}
We note the following runtime considerations.
At its core, \RAWLSNET\, is solving the linear system above.
The two bottlenecks in terms of runtime are (a) the number of constraints and (b) the time it takes to construct the constraints.
For every assignment of \(\justifiedvars\) and \(\sensitivevars\), we have a constraint of the form in Equation~(\ref{eqn:goal}).
While this is exponential in \(|\justifiedvars| + |\sensitivevars|\), we note that in practice both sets are typically small.
To compute the coefficients (i.e., the \(\tmpvar{\cdot}{}\) terms in (\ref{eqn:lhs}) and (\ref{eqn:rhs})), we perform exact inference in the underlying BN.
This is doable in linear time via dynamic programming if the BN is a polytree \cite{russellandnorvig}.
We note that approximate inference (e.g., via particle filtering) is an option for general Bayesian Networks.

\section{Experiments}

We demonstrate the effectiveness of \RAWLSNET\ on the following data: (1) our illustrative college-admissions example (2) another illustrative example regarding campaign financing for elections (3) a synthetic  HR dataset examining employee promotions by IBM, and (4) a real campus recruitment dataset .
All experiments were run on a laptop with 8GB memory and a 1.8 GHz Intel Core i5 processor. We utilized \texttt{cvxpy} to optimize constrained linear systems of equations, \texttt{pgmpy} for Bayesian Network training and inferences, and \texttt{matplotlib} for plotting.

\subsection{Illustrative Example: College Admissions} \label{sec:coll_ex}

We start with our college admissions example, where the BN is specified in Figure~\ref{fig:collegeBN}.  We assign variable names to each of the nodes in the BN. $\talent$ is Talent, $\ses$ is Socioeconomic Status, $\exam$ is Test (Exam) Score, $\college$ is College, and $\goodjob$ is Job. 
With this notation, the justified, sensitive, and other variables are as follows: \justifiedvars = \{\talent\}; \sensitivevars = \{\ses\}; \othervars = \{\exam, \college, \goodjob\}. The control variable is the college admissions policy $\college$ and the query variable is job $\goodjob$. To satisfy FEO, whether someone gets a job must be independent of their SES given their Talent. Therefore, the following equalities must hold: $P(J|T$=Low, $SES$=Low) $\equiv$ $P(J|T$=Low, $SES$=High) and $P(J|T$=High, $SES$=Low) $\equiv$ $P(J|T$=High, $SES$=High). Substituting Equation~\ref{eqn:rhs} into these equations yields a linear system of equations, where the variables are the CPT values for the College node. \RAWLSNET\ solves this linear system.

\begin{figure}
\centering
\begin{subfigure}{\columnwidth}
  \centering
  \includegraphics[width=0.95\linewidth]{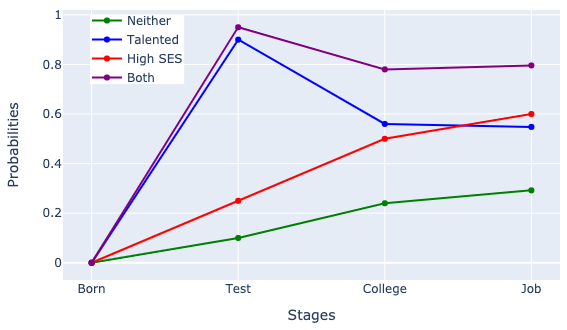}
  \label{fig:college-original}
  \subcaption{Probability of desired outcome before applying \RAWLSNET.}
\end{subfigure}\\
\begin{subfigure}{\columnwidth}
  \centering
  \includegraphics[width=0.95\linewidth]{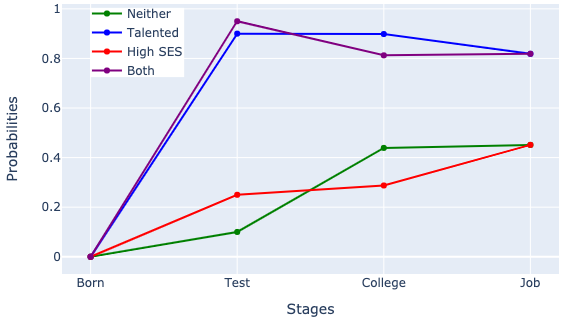}
  \label{fig:sub2}
  \subcaption{Probability of desired outcome after applying \RAWLSNET.}
\end{subfigure}\\
\begin{subfigure}{\columnwidth}
  \centering
  \includegraphics[width=0.95\linewidth]{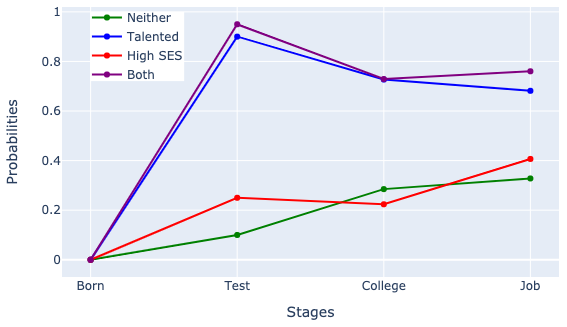}
  \label{fig:sub3}
  \subcaption{Probability of desired outcome after applying \RAWLSNET\ with feasibility constraints.} 
\end{subfigure}
\caption{Visualizations of the outcomes of the synthetic college-admissions BN. The x-axis lists each successive desired outcome and the y-axis represents the proportion of the population that obtained the desired outcome. The population is further broken down on the basis of SES and talent. Except for the college-admissions CPT, the other CPTs were identical. \RAWLSNET\ can achieve FEO with the probability of obtaining a good job being independent of SES.}
\label{fig:college}
\end{figure}

\textbf{Results.}
Figure \ref{fig:college} visualizes the distribution of those receiving good jobs before and after we perform \RAWLSNET. As the figure shows, both distributions share the same testing and hiring policies that have been chosen to encode well-known societal biases, such as the fact that those with high SES backgrounds generally fare better in the hiring process than those of low socioeconomic background.

We highlight the following flexibility of \RAWLSNET. Two of the four CPT values for college admission are fixed, and \RAWLSNET~ is tasked with determining the other two values. As a result, the solution for this particular example can be interpreted as: given the admissions policy for high SES individuals, what policy must be implemented for the low SES applicants to ensure FEO at the job application phase?

The probabilities of college-admissions and job-offers in Figure~\ref{fig:college}(a) agree with our expectations of an unfair world. However, the probabilities after \RAWLSNET has been applied in Figure~\ref{fig:college}(b) fixes this unfairness -- as indicated by the intersection of the blue and purple lines as well as the red and green lines.

We highlight an additional capability of \RAWLSNET: feasibility constraints. In Figure~\ref{fig:college}(c), we show the output of \RAWLSNET\ when the aggregate college-admissions rate is capped at 50\%. When such (linear) constraints are at play, \RAWLSNET\ solves a linear program. 
As expected, the conditional probabilities for Job are no longer equal (as FEO is not achievable under these constraints), but they are much closer to FEO compared to the original distribution (as seen in Figure~\ref{fig:college}(a)). We note that \RAWLSNET\ obtaining probabilities as close as possible to an FEO satisfying distribution is especially helpful when it is used to inform policy makers.  In this example, the college-admissions committee can better understand how the cap on the admissions rate affects their ability to satisfy FEO.

\subsection{Illustrative Example: Campaign Finance}

\begin{figure}
    \centering
    \includegraphics[scale=0.46]{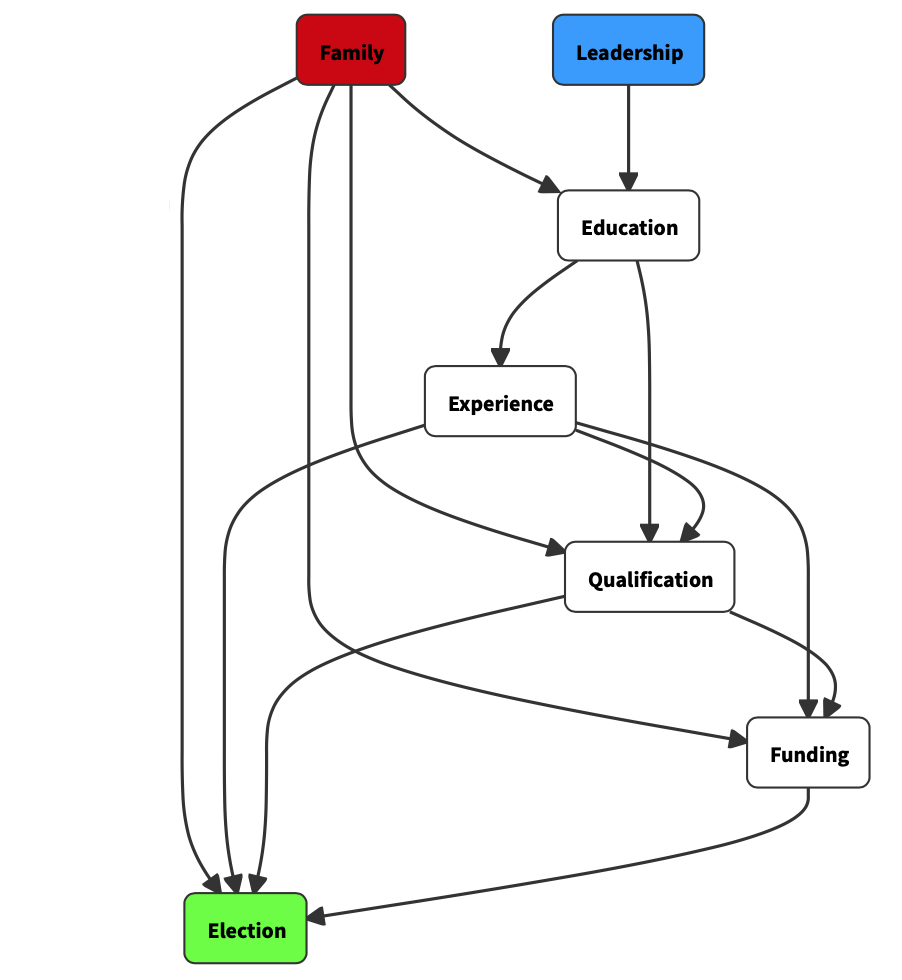}
    \caption{The synthetic campaign finance example sets the probability of being born with $Leadership$ skills as the \textit{justified} variable. The probability of belonging to a $Family$ with political influence is chosen as the \textit{sensitive} variable.  $Funding$ is  the \textit{control} variable. $Election$ is the advantageous social position (i.e., the \textit{target} variable). The variables $Education$, $Experience$, $Qualification$ are \textit{other} variables.}
    \label{fig:campaign-bn}
\end{figure}

To highlight a situation where strict FEO is not satisfiable, we consider the domain of financing a political campaign and show the BN structure for this example in Figure \ref{fig:campaign-bn}. We choose the values for the CPTs  based on our understanding of the real-world dynamics of an election campaign, where we assume that those who come from a family of politicians, have a better chance of obtaining funding and winning elections than those who do not. 
All variables in this example are binary, except the target variable of $Election$ which can take on three values: ``Not Elected'', ``Nominee", and ``Elected".

\textbf{Results.} Table \ref{table:campaign_org} shows the original probabilities for the $Election$ variable given $Leadership$ and $Family$. The goal of \RAWLSNET\ is to modify the CPT for the $Funding$ variable such that for a given $Leadership$ level (i.e., ``Good'' or ``Poor''), the probability of being elected or nominated does not depend on whether or not a person comes from a family with political history.  Given the initial CPTs, \RAWLSNET\ manages to reduce the disparity between the probabilities of winning given leadership and family background (see Table \ref{table:campaign_opt}). However, the conditional probabilities do not satisfy Equation~\ref{eqn:goal}. To solve the system of equations, seven of the eight CPT values for the $Funding$ variable would need to be greater than one. Instead,  \RAWLSNET\ provides valid CPT values that come closest to satisfying Equation~\ref{eqn:goal}, where closest refers to the CPT values that minimize the squared-difference between Equations~\ref{eqn:lhs} and~\ref{eqn:rhs}. 

\begin{table}[!ht]
\centering
\renewcommand{\arraystretch}{1.1}
\begin{tabular}{|c c c c c|} 
    \hline
    Lead. & Family & Not Elected & Nominee & Elected\\ 
    \hline\hline
    Poor & Not Political & 29.20\% & 33.20\% & 37.60\%\\
    Poor & Political & 19.1\% & 8.60\% & 72.30\%\\
    \hline
    Good & Not Political & 27.5\% & 30.20\% & 42.30\%\\
    Good & Political & 17.8\% & 8.10\% & 74.10\%\\
    \hline
\end{tabular}
\caption{Original probability values for $P(Election | Leadership, Family)$ for the Campaign Finance example.}
\label{table:campaign_org}
\end{table}

\begin{table}[!ht]
\centering
\renewcommand{\arraystretch}{1.1}
\begin{tabular}{|c c c c c|} 
    \hline
    Lead. & Family & Not Elected & Nominee & Elected\\ 
    \hline\hline
    Poor & Not Political & 16.34\% & 24.10\% & 59.50\%\\
    Poor & Political & 22.34\% & 11.10\% & 66.60\%\\
    \hline
    Good & Not Political & 15.96\% & 21.40\% & 62.60\%\\
    Good & Political & 21.14\% & 10.70\% & 68.20\%\\
    \hline
\end{tabular}
\caption{Updated probability values for $P(Election | Leadership, Family)$ after using \RAWLSNET\ for the Campaign Finance example.
Note that FEO cannot be satisfied in this case, thus \RAWLSNET\ selects the closest possible distribution.}
\label{table:campaign_opt}
\end{table}

\subsection{Synthetic Data: IBM HR Dataset}
\begin{figure}
    \centering
    \includegraphics[scale=0.5]{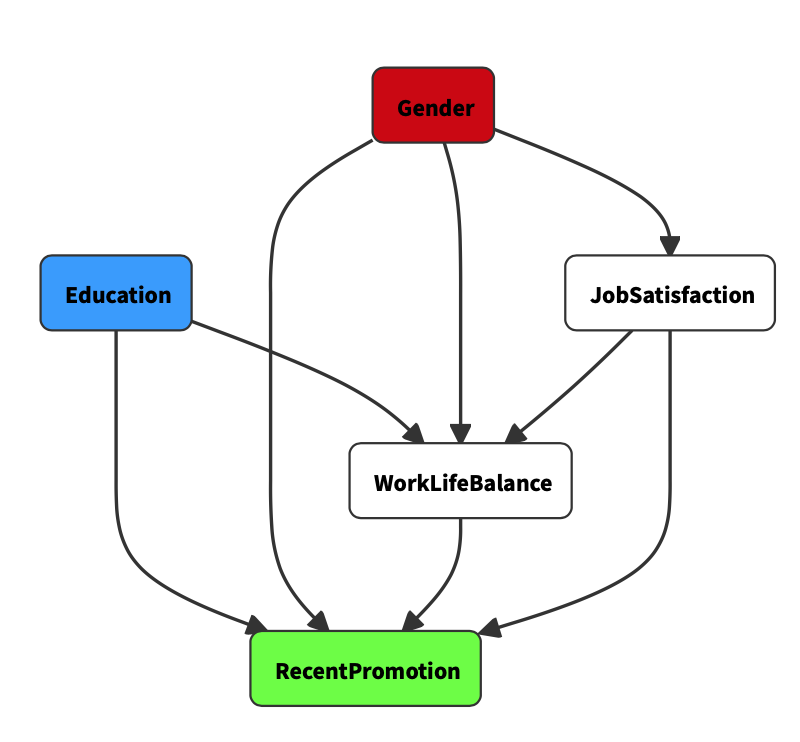}
    \caption{IBM HR Example:  We consider the node $Education$ as a proxy for talent (thus, it is the \textit{justified} variable). $Gender$ is the \textit{sensitive} variable. $WorkLifeBalance$ is the \textit{control} variable since we assume that employers can help alter it. $RecentPromotion$ is  the advantageous social position within the company, and hence is the \textit{target} variable. $JobSatisfaction$ is an \textit{other} variable.}
    \label{fig:ibm_hr}
\end{figure}

The IBM HR Analytics Employee Attrition \& Performance dataset~\cite{ibm_hr} is a synthetic dataset created by IBM to model the factors that lead to employee attrition. We use the dataset to model gender-bias in staff promotions.
$Gender$ is  the \textit{sensitive} variable. $Education$ is a proxy for talent; thus it is the  \textit{justified} variable. $RecentPromotion$ is the advantageous social position (i.e., the \textit{target} variable). We assume that work-life balance (named $WorkLifeBalance$) and $JobSatisfaction$ are the \textit{other} variables, with the $WorkLifeBalance$ also being the \textit{control} variable.  The relationships between the variables are shown in Figure~\ref{fig:ibm_hr}. This data set contains 35 features out of which  4 are related to the FEO task at hand. The variables $Education$ and $JobSatisfaction$ are categorical with 5 and 4 categories, respectively Higher values correspond to greater education and satisfaction. The variable $RecentPromotion$ refers to the number of years since the employee was last promoted. We convert $RecentPromotion$ into a binary variable by thresholding at the median value. The variable $WorkLifeBalance$ is also a categorical variable with 4 categories. We convert it into a binary variable for ease of exposition.

\textbf{Results.} Table~\ref{table:ibm_org} shows the original probabilities of getting a promotion given an education level (Bachelors, Masters, etc.) and gender. We observe slight discrepancies between the promotion probabilities of male and female employees. Table~\ref{table:ibm_opt} shows the results after \RAWLSNET\ was applied to this network, which updated the values of the CPT for the $WorkLifeBalance$ node. The results show that given an education level, the probability of promotions stay the same regardless of gender. 

\begin{table}[ht]
\centering
\renewcommand{\arraystretch}{1.1}
\begin{tabular}{|c c c|} 
    \hline
    Education & Gender & Promotion \\ 
    \hline\hline
    Below College & Male & 36.27\% \\
    Below College & Female & 37.17\% \\
    \hline
    College & Male & 36.27\% \\
    College & Female & 37.17\% \\
    \hline
    Bachelor & Male & 41.79\% \\
    Bachelor & Female & 39.87\% \\
    \hline
    Master & Male & 40.19\% \\
    Master & Female & 40.73\% \\
    \hline
    Doctor & Male & 40.19\% \\
    Doctor & Female & 36.51\% \\
    \hline
\end{tabular}
\caption{Original probability values for $P(RecentPromotion | Education, Gender)$ for IBM HR data.}
\label{table:ibm_org}
\end{table}

\begin{table}[ht]
\centering
\renewcommand{\arraystretch}{1.1}
\begin{tabular}{|c c c|} 
    \hline
    Education & Gender & Promotion \\ 
    \hline\hline
    Below College & Male & 32.61\% \\
    Below College & Female & 32.61\% \\ \hline
    College & Male & 32.61\% \\
    College & Female & 32.61\% \\\hline
    Bachelor & Male & 43.04\% \\
    Bachelor & Female & 43.04\% \\ \hline
    Master & Male & 38.33\% \\
    Master & Female & 38.33\% \\ \hline
    Doctor & Male & 34.80\% \\
    Doctor & Female & 34.80\% \\
    \hline
\end{tabular}
\caption{Updated probability values for $P(RecentPromotion | Education, Gender)$ after using \RAWLSNET\ for IBM HR data.}
\label{table:ibm_opt}
\end{table}

\subsection{Real Data: Campus Recruitment}

The Campus Recruitment Data from Kaggle~\cite{campusrecruitment} contain  information about students in India. It includes their scores from standardized testing, whether or not they got a job at the end of school, and with what salary. While the data set is not primarily designed for making decisions which count as FEO applications, it includes relevant data and is publicly available. It involves at least one decision that affects the distribution of advantageous social positions: whether a student receives a competitive  internship.  We  assume  that  students  receiving internships have a better chance at landing a job later. Thus, we use this decision as an example of an FEO application.

\begin{figure}
    \centering
    \includegraphics[scale=0.45]{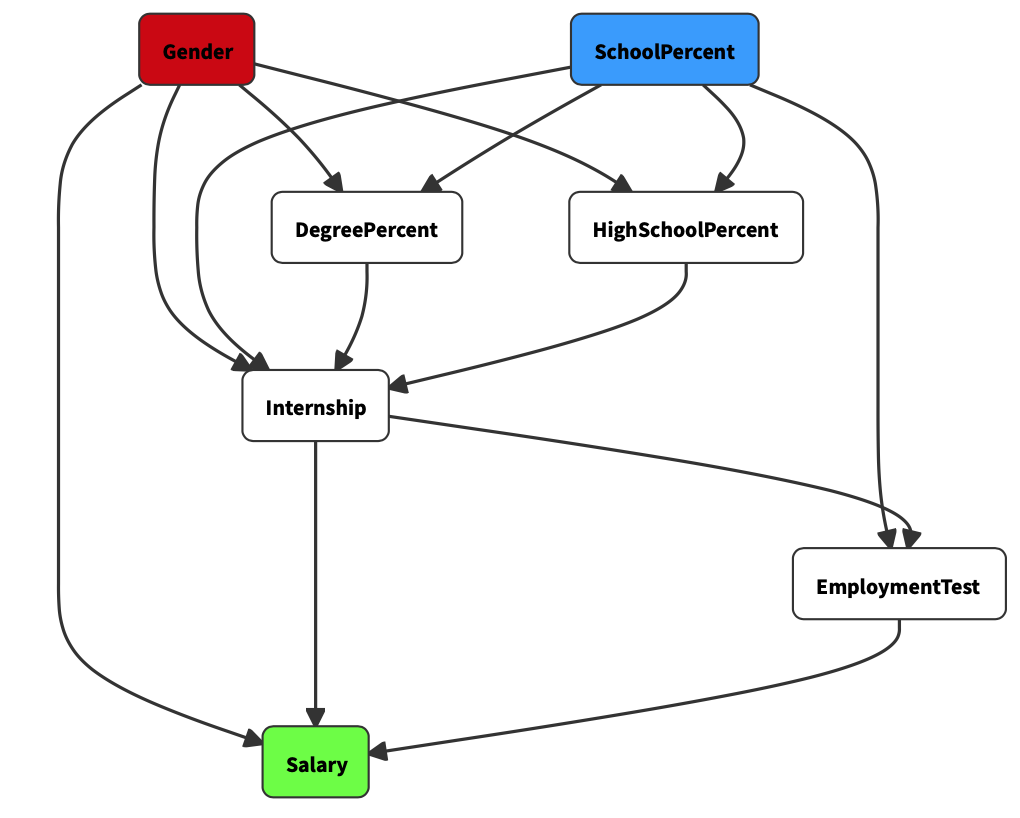}
    \caption{The campus recruitment data looks at the probability of getting a job with a good salary given the innate talent and gender of the applicant. We consider $Gender$ as the \textit{sensitive} variable and $SchoolPercent$ as the \textit{justified} variable. The nodes $HighSchoolPercent$, $DegreePercent$, and $EmploymentTest$ are test scores for high school, undergraduate degree and an employment eligibility test, respectively. They all belong to the \textit{other} variables. $Internship$ looks at distributing internships to students which could help their prospects for a job at a later time. This is the \textit{control} variable. $Salary$ is the advantageous social position and the \textit{target} variable.}
    \label{fig:campus-bn}
\end{figure}

We assume $Gender$ is the \textit{sensitive} variable.  $SchoolPercent$, which represents the earliest standardized test score available for each student, is the \textit{justified} variable. $Internship$ refers to whether the student received a competitive internship and is our \textit{control} variable. $Salary$ is  the advantageous social position (i.e., the \textit{target} variable). The BN shown in Figure~\ref{fig:campus-bn} also includes variables $DegreePercent$ that represents the undergraduate scores, $HighSchoolPercent$ that represents standardized test scores during high school, and $EmploymentTest$ that stands for scores earned in a test that determines eligibility for the job. These variables constitute the \textit{other} variables.\footnote{The variable names have been changed to improve readability. They do not necessarily match the ones in the original dataset.}

The datatset consists of 15 features of which we use 7 that are relevant to the FEO-use case. The variables $SchoolPercent$, $DegreePercent$, $HighSchoolPercent$, $EmploymentTest$ and $Salary$ are continuous variables. We used the median values for each of these variables to convert them to binary variables. The variables $Gender$ and $Internship$ are categorical variables with a cardinality of 2 that contain strings, which were converted to numeric categorical variables. The relationship between these variables are shown in Figure~\ref{fig:campus-bn}. The CPTs for each node were learned from the data using maximum likelihood estimation.

\textbf{Results.} Table~\ref{table:campus_org} shows the original probabilities of getting a good job given $SchoolPercent$ (i.e., the talent proxy) and $Gender$. The table shows that male applicants have a higher probability of getting a good salary as compared to female applicants given the same talents. Table~\ref{table:campus_opt} shows the probabilities of getting a job given talent and gender after \RAWLSNET\ has modified the CPT for the control variable $Internship$. We observe that FEO is satisfied and the probabilities of getting a good salary remain the same irrespective of gender.

\begin{table}[ht]
\centering
\renewcommand{\arraystretch}{1.1}
\begin{tabular}{|c c c|} 
    \hline
    SchoolPercent & Gender & Salary \\ 
    \hline\hline
    Low Score & Male & 50.86\% \\
    Low Score & Female & 30.59\% \\
    \hline
    High Score & Male & 61.87\% \\
    High Score & Female & 33.89\% \\
    \hline
\end{tabular}
\caption{Original probability values for $P(Salary | SchoolPercent, Gender)$ for the Campus Recruitment Data.}
\label{table:campus_org}
\end{table}

\begin{table}[ht]
\centering
\renewcommand{\arraystretch}{1.1}
\begin{tabular}{|c c c|} 
    \hline
    SchoolPercent & Gender & Salary \\ 
    \hline\hline
    Low Score & Male & 43.06\% \\
    Low Score & Female & 43.06\% \\
    \hline
    High Score & Male & 47.76\% \\
    High Score & Female & 47.76\% \\
    \hline
\end{tabular}
\caption{Updated probability values for $P(Salary | SchoolPercent, Gender)$ after using \RAWLSNET\ for the Campus Recruitment Data.}
\label{table:campus_opt}
\end{table}

\section{Related work}

In recent years, Rawls' work has become influential in the algorithmic fairness literature \cite{lundgardMeasuringJusticeMachine2020,barocas-hardt-narayanan,mehrabiSurveyBiasFairness2019}. Some of this work focuses on using the other aspects of Rawls' theory, such as the original position \cite{rawls1971}, to develop novel principles of governance to ensure appropriate transparency, explainability, and fairness \cite{leeProceduralJusticeAlgorithmic2019,wongDemocratizingAlgorithmicFairness2020,graceAITheoryJustice2020}.
Other work has appealed to Rawls' difference principle or prioritarian principles inspired by it \cite{doornRawlsianApproachDistribute2010,lebenRawlsianAlgorithmAutonomous2017,ChenFairness2019}. These works share a general philosophical outlook with our project, as they concern justice as fairness. However, they appeal to different parts of the Rawlsian framework to achieve different goals. Our work is complementary to these others as it implements another aspect of Rawls' theory. Together, these approaches offer the potential for a unified contractualist approach to the ethics of AI. In contrast \citet{lundgardMeasuringJusticeMachine2020}, raises objections to use of Rawls' theory for fair ML, but these objections are less pressing for FEO in particular.

There has also been significant interest in complementary projects in the algorithmic fairness literature which appeal to substantive equality of opportunity principles, including FEO \cite{NIPS2016_6374,dworkFairnessAwareness2012,josephFairnessLearningClassic2016,marrasEqualityLearningOpportunity2020,NIPS2018_7625,kangInFoRMIndividualFairness2020,friedlerImPossibilityFairness2016}. These appeals are used to justify various specific fairness metrics, and to adjudicate disputes between these metrics. Work in this literature appeals to Rawls' FEO and similar substantive equality of opportunity principles, in particular those discussed and formalized by John Roemer~\cite{roemerEqualityOpportunity2009a,roemerEqualityOpportunity2015}. The primary difference between our project and these others is that they are concerned with determining appropriate metrics of fairness. In other words, they are looking to measure the degree to which particular uses of ML algorithms count as fair. They use these metrics to evaluate and mitigate bias in ML. Many of those who appeal to FEO do so in order to argue for one or another fairness metric as better than alternatives, or at least better for a particular type of circumstance. For instance, Binns argues that FEO considerations justify appeals to group fairness metrics in certain contexts and individual fairness metrics in other contexts \cite{binnsApparentConflictIndividual2020}. Loi et al.~argue that a modified, generalized version of FEO justifies using two different fairness metrics, sufficiency and separability, in distinct contexts \cite{loiPhilosophicalTheoryFairness2019a}. Heidari et al.~similarly argue that various notions of algorithmic fairness can be justified as special cases of a substantive EO principles like FEO \cite{heidariMoralFrameworkUnderstanding2019}.

Our project offers a useful addition to  these other substantive equality of opportunity approaches. \RAWLSNET\ is novel in that it models interventions on decisions. It computes which interventions will obtain FEO (or promote it as much as possible). Thus, the goal of our project is significantly different than the goal of those in the fairness metric literature. In addition, \RAWLSNET's output distribution can be sampled to generate new aspirational data, or it can be used to inform decision-makers. It is less concerned with evaluating the performance of ML algorithms, though it might be useful for that purpose, as we hope to explore in future work. 

\subsection{Training Fair Graphical Models}
Previous efforts in training fair BNs achieve fairness by either re-training the BN with re-labeled data \cite{mancuhan_combating_2014} or imposing fairness constraints during parameter learning \cite{Choi_2020}.
 Our approach instead directly modifies the appropriate conditional probability values without changing the structure of the BN. Unlike past approaches \cite{CardosoFramework2019}, our goal is to generate fair data distributions, which can subsequently be used for sampling aspirational data or guiding policy decisions.

Previous attempts have been made to generate fair data with other models, such as GANs \cite{fairgan}. However, our approach is unique in basing our definition of fairness on Rawls and in producing aspirational data distributions. As such, we are able to provide clearer guidance on when aspirational data generated through \RAWLSNET\ should and should not be used.

\section{Discussion}
\label{sec:uses}
We presented a method, called \RAWLSNET, which determines how a BN model of an FEO application must be altered in order to satisfy FEO. This system offers the ability to model circumstances of ideal fairness in order to generate distributions over aspirational (FEO) data. This aspirational data distribution can be used by researchers to promote fair ML by sampling from it to discover and avoid pitfalls in ML algorithms, which  can lead to unfairness despite unbiased data. \RAWLSNET\ can also be used to offer advice to decision-makers seeking to promote FEO. However, caution must be exercised when using \RAWLSNET\ for this purpose. The accuracy of the results of \RAWLSNET\ depend on the accuracy of the assignments of the  variables in the data to the appropriate justified, sensitive, and other categories. We highlight this issue, in particular, for using \RAWLSNET\ for policy advice. Talent is not directly observable. Talent refers to an individual's innate, intrinsic features which partially determine his/her capability for succeeding and excelling in a social position. These features are only indirectly observable. Moreover, the actual evidence we have regarding innate talent will often be confounded by the complexity of the social systems which impact individual education and development. For instance, a good proxy for talent may be early standardized testing. However, a student's test scores will also certainly be influenced by their early home life, which is in turn influenced by factors like socioeconomic status. Thus, the very bias we are attempting to eliminate may creep into the proxy we use to evaluate talent.

Our use of the campus recruitment data \cite{campusrecruitment} illustrates some of the problems for using our system for policy advice. In order to make use of a public data set we used an imperfect proxy for talent. We assigned the ``SchoolPercent'' feature, which refers to standardized test scores, to the justified variable category on the assumption that it was the best available proxy for talent. However, in this case the proxy will certainly be imperfect, as sensitive features are likely to impact this test score. \RAWLSNET\ would do a better job at advising policy decisions to promote FEO with a better proxy for talent. As it stands, the recommendations of \RAWLSNET\, for this example may help to promote FEO, but they will not be able to guarantee it without a better, less biased proxy for talent. In future research, we intend to pair \RAWLSNET\ with methods that infer the distribution of unobserved features such as innate talent.

\RAWLSNET\ cannot, by itself, ensure fairness in decision-making. However, it does provide useful information for domain experts, who might use it to ask questions about what it would take to satisfy FEO in their own decisions. It is not a system that is ready to guide any user in making decisions based on any data set. However, it can be a useful method in the hands of domain experts who can make appropriate judgments about what FEO requires, and about how to apply \RAWLSNET. To illustrate this, imagine a group of political scientists developing policy proposals for ameliorating the effect of SES on standardized test scores. They have done a number of empirical studies to gather data in different parts of the country. Using a variety of ML methods, they create a set of Bayesian Networks that model different possible ways the world could be, consistent with their data. They are not sure which one is the most accurate. These scientists could use \RAWLSNET\ to find the best policy to promote FEO in each one of these BN models. If the one policy for giving out extra training is successful in promoting FEO in all their BN models, this could provide support for implementing that policy. Alternatively, it might be that a policy works well in some but fails disastrously in others. This would then be a sign that more research is needed before implementing the policy. In this way, \RAWLSNET\ provides useful information, despite the fact that one cannot simply deploy it without domain knowledge.

\bibliography{rawlsnet}

\end{document}